\documentclass[twocolumn,aps,prb]{revtex4-2}
\usepackage[latin9]{inputenc}
\usepackage{mathtools}
\usepackage{amsmath}
\usepackage{amsfonts}
\usepackage{mathrsfs}
\usepackage{graphicx}
\usepackage{esint}
\usepackage[unicode=true, bookmarks=false, breaklinks=false, pdfborder={0 0 1}, backref=false, colorlinks=false]{hyperref}
\usepackage{url}
\usepackage[dvipsnames]{xcolor}
\usepackage{hyperref}
\usepackage{xfrac}

\usepackage[draft]{changes}

\begin{document}

\newcommand{\const}{\text{const}}
\newcommand{\dd}{\partial}
\newcommand{\<}{\langle}
\renewcommand{\>}{\rangle}
\renewcommand{\(}{\left(}
\renewcommand{\)}{\right)}
\renewcommand{\[}{\left[}
\renewcommand{\]}{\right]}
\newcommand{\lw}{\left\{}
\newcommand{\pw}{\right\}}

\newcommand{\opb}{\hat{b}}
\newcommand{\opbd}{\hat{b}^{\dag}}
\newcommand{\opn}{\hat{n}}

\newcommand{\vc}{\overline{v}}
\newcommand{\Vc}{\overline{V}}
\newcommand{\uc}{\overline{u}}
\newcommand{\Uc}{\overline{U}}
\newcommand{\cc}{\overline{c}}
\newcommand{\cPsi}{\overline{\Psi}}
\newcommand{\cA}{\overline{A}}
\newcommand{\cB}{\overline{B}}
\newcommand{\cC}{\overline{C}}
\newcommand{\np}{\tilde{n}}
\newcommand{\nsqrt}{\overline{n}_0}
\newcommand{\tPsi}{\tilde{\Psi}}

\newcommand{\K}{\mathbf{k}}
\newcommand{\q}{\mathbf{q}}
\renewcommand{\r}{\mathbf{r}} 

\newcommand{\remark}{\textcolor{WildStrawberry}}


\title{Excitation spectra of strongly interacting bosons in the flat-band Lieb lattice}

\author{B. Grygiel}
\author{K. Patucha}
\email[Corresponding author: ]{k.patucha@intibs.pl}
\affiliation{Institute of Low Temperature and Structure Research, Polish Academy of Sciences, Ok\'olna 2, 50-422 Wroc\l{}aw, Poland}

\begin{abstract}
The strongly correlated bosons in flat band systems are an excellent platform to study a wide range of quantum phenomena. Such systems can be realized in optical lattices filled with ultracold atomic gases. In this paper we study the Bose-Hubbard model in the Lieb lattice by means of the time-dependent Gutzwiller mean-field approach. We find that in the Mott insulator phase the excitation modes are gapped and display purely particle or purely hole character, while in the superfluid phase the excitation spectrum is gapless. The geometry of the Lieb lattice leads to a non-uniform order parameter and non-uniform oscillation energy in the ground state. This results in additional anti-crossings between dispersive bands in the excitation spectra, while the flat bands remain insensitive to this effect. We analyze the oscillations of the order parameter on the sublattices as well as the particle-hole character of the excitations. For certain model parameters we find simultaneous pure phase and pure amplitude oscillations within the same mode, separated between the sublattices. Also, we propose a simple method to differentiate between the hole- and particle superfluid regions in the Lieb lattice by in-situ measurement of the atom population on the sublattices.
\end{abstract}

\maketitle


\section{Introduction}

The flat band systems exhibit a lot of interesting many-body phenomena as a result of strong spatial localization of the states in the flat band. As there is no momentum dependence of the single-particle energy in the flat band, the influence of the interaction and disorder is enhanced. Flat bands naturally occur in some lattices due to destructive interference of the single particle wavefunction~\cite{vidal_aharonov-bohm_1998,bergman_band_2008,liu_exotic_2014,mukherjee_observation_2015-1,leykam_artificial_2018,nunes_flat-band_2020,peri_fragile_2021,kumar_flat-band-induced_2021} with well known examples being the Lieb lattice and the dice ($\mathcal{T}_3$) lattice. They are so called decorated lattices with one type of sites (hub) having a larger number of the nearest neighbors than the other sites (rim, located on the links between the hub sites)~\cite{leykam_artificial_2018}. In these systems the flat band is protected by the lattice symmetry~\cite{leykam_artificial_2018}.

The ferromagnetic, superconducting, and topological properties of the Lieb lattice based models were studied extensively~\cite{lieb_two_1989,noda_ferromagnetism_2009,goldman_topological_2011,iglovikov_superconducting_2014,noda_flat-band_2014,noda_magnetism_2015,tsai_interaction-driven_2015,palumbo_two-dimensional_2015}, especially in the context of $\text{CuO}_2$ planes in high-$T_c$ superconductors~\cite{emery_theory_1987,varma_charge_1987,varma_non-fermi-liquid_1997}. It has been also shown that the quantum metric of the flat band gives a major contribution to the superfluid weight~\cite{julku_geometric_2016,torma_quantum_2018} and dc conductivity~\cite{mitscherling_bound_2022}. Moreover, at M point of the Brillouin zone (BZ) the flat band intersects with a Dirac cone, which makes the Lieb lattice a good model to capture the physics of twisted bilayer graphene~\cite{hu_geometric_2019,julku_superfluid_2020,xie_topology-bounded_2020,salamon_simulating_2020}. This feature also enables to study a crossover between massive and massless pseudospin-1 Dirac fermions~\cite{shen_single_2010}.

Bosonic systems in lattices with flat bands also exhibit interesting behavior e.g., the condensation in the flat band is supressed and is driven by the interaction energy~\cite{huber_bose_2010}. Geometry of the lattice has a direct influence on the order parameter of the Bose-Hubbard (BH) model in the dice lattice, making it spatially non-uniform with stronger phase localization on the hub sites~\cite{rizzi_phase_2006}. Moreover, in the presence of the magnetic field with the $\sfrac{1}{2}$-flux a new insulating, commensurate phase emerges due to boson localization in the Aharonov-Bohm cages~\cite{rizzi_phase_2006,moller_correlated_2012}. For bosons in the $p$-band Lieb lattice it was shown that the excitations exhibit anomalous Hall effect~\cite{di_liberto_topological_2016}.

Although the physics of the flat band lattices was studied theoretically, only the recent technological advancements allowed of their experimental realization. The Lieb lattice was created in the optical lattice systems with fermionic~\cite{shen_single_2010,taie_spatial_2020} as well as bosonic~\cite{taie_coherent_2015,ozawa_interaction-driven_2017} ultracold atoms. These systems are highly tunable and allow of a population of arbitrary lattice sites and a transfer of the atoms into higher energy bands. There are also reports of realizations of the Lieb lattice in optical cavity systems~\cite{baboux_bosonic_2016} as well as photonic~\cite{guzman-silva_experimental_2014,mukherjee_observation_2015,vicencio_observation_2015,diebel_conical_2016} and electronic~\cite{slot_experimental_2017} lattices. 

In this paper our goal is to analyze the excitation spectra of the Bose-Hubbard model in the Lieb lattice. The BH model was first proposed by Berlin and Kac~\cite{berlin_spherical_1952} to describe strongly interacting bosons in a lattice. The model exhibits quantum phase transition between the superfluid (SF) and Mott insulator (MI) states, which is driven by the competition between the kinetic and the on-site interaction energy~\cite{fisher_boson_1989}. In the MI phase the excitation modes are gapped and exhibit either particle or hole character, while in the SF phase the first excited mode is a gapless Goldstone mode and the higher modes are gapped. For certain parameters the model can also exhibit a massive Higgs mode~\cite{di_liberto_particle-hole_2018}.

The excitation spectra of the BH model were studied by means of various techniques, e.g. the random phase approximation~\cite{sheshadri_superfluid_1993,van_oosten_inelastic_2005,gerbier_interference_2005,
menotti_spectral_2008,hazzard_many-body_2010}, Schwinger-boson mean-field approach~\cite{huber_dynamical_2007}, strong-coupling method~\cite{elstner_dynamics_1999,sengupta_mott-insulator----superfluid_2005}, variational cluster approach~\cite{knap_spectral_2010}, quantum rotor approach~\cite{zaleski_momentum-resolved_2012,zaleski_optical_2012}, and dynamical mean field theory~\cite{panas_numerical_2015}. The properties of the Higgs and Goldstone modes in the BH model were considered in the context of the entanglement entropy~\cite{frerot_entanglement_2016}, topological character of the bands~\cite{wang_topological_2021}, and the impact on the impurity behavior~\cite{caleffi_impurity_2021}. 

Inspired by M. Di Liberto~\emph{et al.}~\cite{di_liberto_particle-hole_2018} we focus on the amplitude and phase character of the excitations, which are related to the Higgs and Goldstone modes, respectively. We also analyze the particle and hole character of the excitations, which can be relevant for finding the dark soliton solution~\cite{krutitsky_dark_2010} and vortex configuration~\cite{wu_vortex_2004}. Moreover, we propose an easy and experimentally accessible method for distinguishing between the particle- and hole superfluid, which employs unique features of the Lieb lattice.

To solve the BH model we use the time-dependent Gutzwiller mean-field approach as presented in~\cite{di_liberto_particle-hole_2018,krutitsky_excitation_2011,krutitsky_ultracold_2016}. Although this method neglects the correlations between the lattice sites, it takes into account the on-site fluctuations and thus allows of a good description of the SF-MI phase transition~\cite{krutitsky_excitation_2011,krutitsky_ultracold_2016}. The Gutzwiller method was successfully used to describe the Higgs mode in the low frequency regime in the two-dimensional optical lattice experiment~\cite{endres_higgs_2012}. The excitation spectra of the BH model on the Lieb lattice exhibit features similar to simple lattices e.g., energy gap in the MI phase, Goldstone mode in the SF phase. However, the unique geometry of the Lieb lattice leads to a non-uniform SF groundstate as well as introduces modifications to the excitation spectra.

The remainder of the paper is structured as follows. In Sec.~\ref{sec:model} we introduce the BH model in the Lieb lattice and present the time-dependent Gutzwiller approach in the case of the considered multiband system. The details of the calculations are relegated to the appendix. Sec.~\ref{sec:exc-spectra} contains the analysis of the excitation spectra. The results are summarized in Sec.~\ref{sec:summary}.


\section{Strongly-interacting bosons in the Lieb lattice}
\label{sec:model}

\begin{figure}[h]
\begin{centering}
\includegraphics[width=\columnwidth]{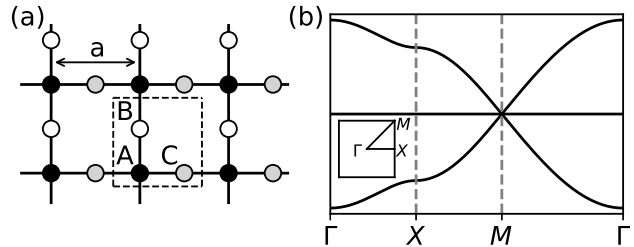} 
\par\end{centering}
\vspace{-3mm}
\caption{(a)~Lieb lattice. The unit cell, marked with dashed line, includes three lattice sites $A$ (hub), $B$ and $C$ (rim). (b)~The energy spectrum of the Lieb lattice along the high symmetry lines~$\Gamma-X-M-\Gamma$. A bulk flat band intersects at the $M$ point with a Dirac cone created by the other two bands. A small inset presents the Brillouin zone with high-symmetry lines.}
\label{fig:lieb-lattice} 
\end{figure}

Here we consider the Bose-Hubbard model on the Lieb lattice. The Lieb lattice is a line-centered square lattice with the elementary cells consisting of three sites: hub sites~$A$ and rim sites~$B$ and~$C$ [see Fig.~\ref{fig:lieb-lattice}(a)]. The tight binding model on the Lieb lattice exhibits a flat band in the entire Brillouin zone with a vanishing wavefuntion amplitude on the $A$ sublattice. At the $M$ point of the BZ the first and the third band create a Dirac cone, which intersects with the flat band [see Fig.~\ref{fig:lieb-lattice}(b)].

\subsection{Hamiltonian}

The Hamiltonian of the system is given in the second-quantization formalism
\begin{align}
\hat{H} &= -J \sum_{l=1}^N \( \opbd_{l,B} \, \opb_{l,A} + \opbd_{l+e_y,A} \, \opb_{l,B} + H.c. \right. \nonumber \\
&+\left.  \opbd_{l,C} \, \opb_{l,A} + \opbd_{l+e_x,A} \, \opb_{l,C} + H.c.\) \nonumber \\
&+ \sum_{l=1}^N \sum_{X=\{A,B,C\}} \[ \frac{U}{2} \, \opn_{l,X} \( \opn_{l,X} -1\) - \mu \, \opn_{l,X} \],
\label{eq:H-BH-Lieb}
\end{align}
where~$J$ denotes the tunneling coefficient between neighboring sites, $U>0$ is the repulsive interaction energy between the atoms on the same lattice site, and~$\mu$ is the chemical potential, which determines the number of particles in the system. The index~$l=(l_x,l_y)$ runs over the elementary cells, while the position of the atom within the elementary cell is given by the index~$X=A,B,C$. The index~$l+e_{x(y)}$ denotes the neighboring elementary cell shifted by a lattice constant~'$\mathrm{a}$' in the $x$($y$) direction. The operator~$\opb_{l,X}$ ($\opbd_{l,X}$) annihilates (creates) an atom on the $X$ lattice site in the~$l$-th elementary cell, while the number operator~$\opn_{l,X}$ counts the particle therein.

\subsection{Time-dependent Gutzwiller ansatz}

To solve the BH model on the Lieb lattice we employ the time-dependent Gutzwiller ansatz~\cite{krutitsky_excitation_2011,krutitsky_ultracold_2016,di_liberto_particle-hole_2018}, in which the eigenstates of the Hamiltonian~(\ref{eq:H-BH-Lieb}) can be represented as a tensor product of the local states
\begin{align}
|\phi \> = \bigotimes_l |\varphi_A \>_l \otimes |\varphi_B \>_l \otimes |\varphi_C \>_l,
\end{align}
where $|\varphi_X \>_l = \sum_{n} c^X_{l,n}(t) \, |n\>_l$. To ensure the normalization of the state, the coefficients~$c^X_{l,n}(t)$ satisfy the following relation
\begin{align}
\sum_{n=0}^\infty |c^X_{l,n}(t)|^2 = 1.
\end{align}
The superfluid order parameter~$\psi_l^X$ is defined as the average of the annihilation operator~$\opb_l^X$ 
\begin{align}
\psi_l^X(t) = \< \opb_{l,X}(t) \> = \sum_{n=1}^\infty \sqrt{n} \, \cc^X_{l,n-1}(t) \, c^X_{l,n}(t).
\end{align}
Minimization of the Lagrangian of the system 
\begin{align}
\mathcal{L} = \sum_l {}_l \<\phi| ( i\hbar \dd_t - \hat{H} ) |\phi\>_l 
\end{align}
where~$\dd_t = \dd/\dd t$, leads to a system of Gutzwiller equations for the coefficients~$c^X_{l,n}$
\begin{align}
i\hbar \, \dd_t \, c^X_{l,n} &= E_n \, c^X_{l,n} \nonumber \\
&- J \, \Psi^X_l \, \sqrt{n} \, c^X_{l,n-1} - J \, \cPsi^X_l \, \sqrt{n+1} \, c^X_{l,n+1},
\label{eq:Gutzwiller_eqs}
\end{align}
where the on-site energy~$E_n = \frac{U}{2} n(n-1) - \mu n$. For the sake of brevity we have omitted the time dependence of the coefficients~$c^X_{l,n}$, parameters~$\Psi^X_l$ and their complex conjugation~$\cPsi^X_l$. Parameters~$\Psi^X_l$ are defined for each lattice site A, B, and C within the elementary cell and reflect the influence of the superfluid order parameter~$\psi$ from the neighboring sites. The symmetry of the Lieb lattice prevents a direct hopping between B and C sites (we assume there is no next nearest neighbor hopping), thus the parameters $\Psi^X_l$ are given as 
\begin{subequations}
\label{eq:effective_param}
\begin{align}
\Psi^A_l &= \psi^B_l + \psi^B_{l-e_y} + \psi^C_l + \psi^C_{l-e_x}, \\
\Psi^B_l &= \psi^A_l + \psi^A_{l+e_y}, \\
\Psi^C_l &= \psi^A_l + \psi^A_{l+e_x}.
\end{align}
\end{subequations}
To solve the Gutzwiller equations~(\ref{eq:Gutzwiller_eqs}), we assume a small perturbation to the ground state solution~$c^X_n$:
\begin{align}
c^X_{l,n}(t) = \[ c^X_{n} + \delta c^X_{l,n}(t) \] e^{-i\omega^X_0 t}.
\label{eq:coeff}
\end{align}
In the case of the Lieb lattice the ground state solution~$c^X_{n}$ depends on the position within the elementary cell, but does not depend on the position of the elementary cell itself. 

\subsubsection*{$0^\text{th}$ order expansion}

Inserting the coefficients~(\ref{eq:coeff}) into the system of Gutzwiller equations~(\ref{eq:Gutzwiller_eqs}) and keeping only zero-order terms in~$\delta c$ allows us to obtain a system of self-consistent equations for the ground state. 

The stationary value of the order parameter is defined for each sublattice
\begin{align}
\psi_X = \sum_{n=0}^\infty \sqrt{n} \, c_{n-1}^X \, c_n^X.
\end{align}
In the MI phase the coefficients
\begin{align}
c_n^X = \delta_{n,n_0},
\label{eq:coeff-MI-GS}
\end{align}
where~$n_0 = \lceil \mu/U \rceil$ is the average number of particles per site and is identical for both sublattices. The function~$ \lceil x \rceil$ denotes the smallest integer greater than~$x$. From Eq.~(\ref{eq:coeff-MI-GS}) it follows that the order parameter~$\psi_X$ vanishes and the MI state is incompressible. The oscillation energies~$\omega_0$ are equal on both sublattices.

\begin{figure}[h]
\begin{centering}
\includegraphics[width=\columnwidth]{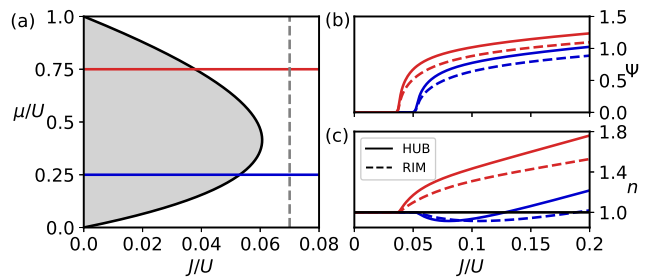} 
\par\end{centering}
\vspace{-3mm}
\caption{(Color online) (a) Phase diagram of the Bose-Hubbard model in the Lieb lattice. The grey area is the first Mott insulator lobe with the average on-site density~$\<n\> = 1$. Beyond the black line the system is in the SF phase. (b) The order parameter on the hub (solid lines) and the rim (dashed lines) sublattice. The blue (lower) pair of lines corresponds to the blue (lower) line in the phase diagram (a) at constant~$\mu/U = 0.25$. The red (upper) pair corresponds to the red (upper) line in the phase diagram at constant~$\mu/U = 0.75$. (c) The average density on the hub (solid lines) and the rim (dashed lines) sublattice. Color labeling is the same as in (b).}
\label{fig:phase-diagram} 
\end{figure}

In the SF phase the coefficients~$c_n^X$ need to be evaluated in a self-consistent manner from Eqs.~(\ref{eq:Gutzwiller_eqs})~\cite{fisher_boson_1989,krutitsky_ultracold_2016}. The phase diagram of the SF-MI transition is presented in Fig.~\ref{fig:phase-diagram}(a). The order parameter~$\psi_X$ takes greater value on the hub sublattice [see Fig.~\ref{fig:phase-diagram}(b)]. Moreover, the ground state oscillation energy for the hub sublattice~$\omega_0^A$ is lower than for the rim sublattice~$\omega_0^B = \omega_0^C$. These two features reflect a stronger phase localization on the hub sites~\cite{rizzi_phase_2006}.

\subsubsection*{$1^\text{st}$ order expansion}

Keeping only the linear terms in~$\delta c$ in Eq.~(\ref{eq:Gutzwiller_eqs}) and employing the following form of the perturbation
\begin{align}
\delta c^X_{l,n}(t) = u^X_{\K,n} \, e^{i \(\K \cdot \r_l^X - \omega_\K t\)} + \vc^X_{\K,n} \, e^{-i \(\K \cdot \r_l^X - \omega_\K t\)}
\label{eq:perturbation}
\end{align}
allows us to obtain a set of equations for the coefficients~$u_{\K,n}^X$ and~$v_{\K,n}^X$. We represent these equations in the matrix form
\begin{align}
\hbar \omega_\K 
\begin{bmatrix}
\vec{u}_\K \\
\vec{v}_\K
\end{bmatrix} = 
\begin{bmatrix}
M_\K & N_\K \\
-N_\K & -M_\K
\end{bmatrix}
\begin{bmatrix}
\vec{u}_\K \\
\vec{v}_\K
\end{bmatrix}
\label{eq:Bogoliubov-eqs}
\end{align}
by introducing vectors~$\begin{bmatrix} \vec{u}^\top_\K, \vec{v}^\top_\K
\end{bmatrix}$, where $\vec{u}^\top_\K = \begin{bmatrix} u^A_{\K,0}, u^B_{\K,0}, u^C_{\K,0}, u^A_{\K,1}, u^B_{\K,1}, u^C_{\K,1} \dots \end{bmatrix}$ and analogoulsy for~$\vec{v}^\top_\K$. The explicit form of the $M_\K$ and $N_\K$ matrices are presented in the Appendix along with other details of the derivation. 

The perturbation~$\delta c^X_{l,n}(t)$~(\ref{eq:perturbation}) leads to the fluctuation of the order parameter around its ground state value~$\psi_X$. For each mode~$\lambda$ it can be expressed as a plane wave 
\begin{align}
\delta\psi_{l,\lambda}^X(t) = U_{\K,\lambda}^X e^{i \(\K\cdot r_l^X - \omega_{\K, \lambda} t\)} + V_{\K,\lambda}^X e^{-i \(\K\cdot r_l^X - \omega_{\K, \lambda} t\)},
\label{eq:order-param-fluct}
\end{align}
where 
\begin{align}
U_{\K,\lambda}^X &= \sum_{n=0}^\infty c_n^X \( \sqrt{n+1} \, u^X_{\K,\lambda,n+1} + \sqrt{n} \, v^X_{\K,\lambda,n-1} \), \\
V_{\K,\lambda}^X &= \sum_{n=0}^\infty c_n^X \( \sqrt{n} \, u^X_{\K,\lambda,n-1} + \sqrt{n+1} \, v^X_{\K,\lambda,n+1} \).
\end{align}
The~$U^X_{\K,\lambda}$ and~$V^X_{\K,\lambda}$ coefficients are the starting point for our further calculations. As shown in Ref.~\cite{krutitsky_ultracold_2016} they can be used to calculate the perturbation of the total density and of the condensate density. Here we utilize them to determine the phase-amplitude character (the flatness parameter) and the particle-hole character of the modes. 

\section{Excitation spectra}
\label{sec:exc-spectra}

The system of Bogoliubov-like equations~(\ref{eq:Bogoliubov-eqs}) is greatly simplified in the MI phase, as it couples only coefficients with~$n = n_0-1$ and~$n_0+1$. This results in a set of 12 equations for~$u_{\K,n_0\pm 1}^X$ and~$v_{\K,n_0\pm 1}^X$ with~$X=A, B, C$. In the SF phase the full system of equations~(\ref{eq:Bogoliubov-eqs}) must be diagonalized, however for the practical purposes the basis can be truncated at a given particle number~$n_\text{max}$. Since all the calculation were performed for~$0 < \mu/U < 1$, where the average particle number per site does not exceed~$\<n\>=2$, we have chosen~$n_\text{max}=5$.

\begin{figure}[h]
\begin{centering}
\includegraphics[width=\columnwidth]{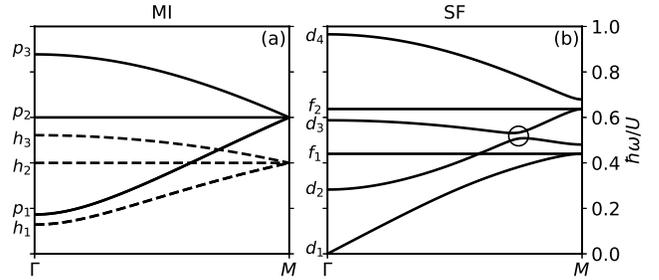} 
\par\end{centering}
\vspace{-3mm}
\caption{The excitation spectra of the strongly interacting bosons in the Lieb lattice in the MI (a) and SF (b) phases. BH model parameters: $\mu/U=0.4$, $t/U=0.055$ in the MI phase, and $t/U=0.07$ in the SF phase. (a) The lower set of bands (dashed lines) labeled with~$h_\lambda$, where $\lambda = 1,2,3$ has a pure hole character, while the upper (solid line) labeled with~$p_\lambda$ -- pure particle character. (b) The excitation bands in SF are grouped in two categories: dispersive $d$ and flat $f$ bands. The circle indicates the anti-crossing between the second and third dispersive bands.}
\label{fig:exc-spectra} 
\end{figure}

The excitation spectra in the MI and SF phases are presented in Fig.~\ref{fig:exc-spectra}(a) and (b), respectively, for the momentum~$\K$ along $\Gamma - M$ line and for the chemical potential~$\mu/U=0.4$ near the particle-hole symmetry line. 

In the Mott insulator phase the spectrum is gapped similarly as for regular, simple lattices. It consists of six lowest bands, grouped into two sets of three, and higher dispersionless bands (not shown). The bands in each set correspond to the energy bands of the non-interacting particles on the Lieb lattice. Always one set exhibits purely hole character, while the other purely particle character. Depending on the BH model parameters we can observe crossing between bands from different sets [Fig.~\ref{fig:exc-spectra}(a)].

\begin{figure}[h]
\begin{centering}
\includegraphics[width=\columnwidth]{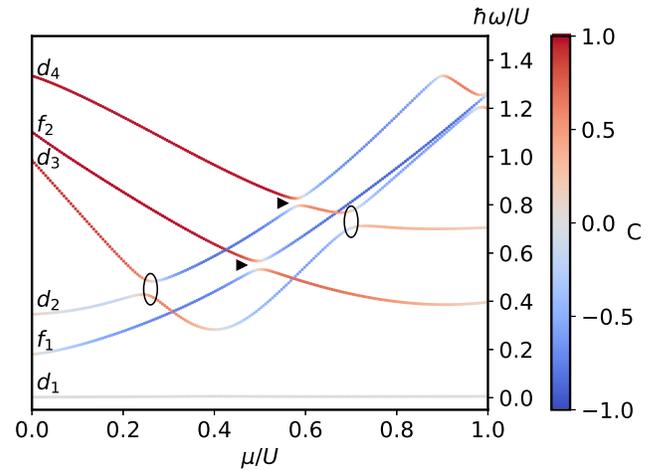} 
\par\end{centering}
\vspace{-3mm}
\caption{(Color online) The first six excitation energy bands for small wavevector~$\K = (\pi/100, \pi/100)$ as a function of the chemical potential~$\mu/U$ for the hopping integral~$J/U = 0.07$ (along the grey dashed line in Fig.~\ref{fig:phase-diagram}(a). The color coding indicates the particle-hole character~$C$ of the excitation. Band labelling is the same as in Fig.~\ref{fig:exc-spectra}(b). Black triangles indicate the regular anti-crossing points, the ellipsis -- the additional anti-crossing points resulting from the non-uniform ground state. The later correspond to the anti-crossing point marked with a circle in Fig.~\ref{fig:exc-spectra}(b) but for different values of the wavevector~$k$ and chemical potential~$\mu/U$.}
\label{fig:exc-spectra-smallk} 
\end{figure}

In the superfluid phase the spectrum is gapless and the lowest mode (Goldstone mode) exhibits linear dispersion near $\K=0$. The seventh and higher excitation bands display very weak dependence on momentum and thus are not shown. In contrast to the MI phase, the degeneracy of the bands at the $M$ point is partially lifted, which results in a gap opening within each set [see Fig.~\ref{fig:exc-spectra}(b)]. This behavior can be attributed to different oscillation energies~$\hbar\omega_0^X$ on the hub and rim sublattices. These energies enter the Bogoliubov-like equations~(\ref{eq:u-Bogoliubov-like}) and~(\ref{eq:v-Bogoliubov-like}) diagonally and thus their difference can be interpreted as an effective potential on the hub sublattice~$\Delta_\text{eff} = \hbar(\omega_0^B - \omega_0^A)$~\cite{lundh_dynamic_2006,bardyn_chiral_2016}. This potential is always positive as $\omega_0^A < \omega_0^B$. Moreover, it is not scaled by the on-site particle number. Analogous influence on the excitation energies was observed for the Haldane-Bose-Hubbard model, where topological transition was shifted by an effective on-site potential resulting from the inequivalence of the sublattices~\cite{furukawa_excitation_2015}. Because the excitation modes originate from the energy bands of the non-interacting system, the influence of an on-site potential~$\Delta$ on the hub sublattice can be understood by considering the non-interacting system. The non-zero potential opens an energy gap at the~$M$ point of the BZ. For~$\Delta>0$ the third band [see Fig.~\ref{fig:lieb-lattice}(b)] separates from the remaining bands, while the first and the second (flat) bands are still degenerate. For~$\Delta <0$ the first band becomes separated and the second and the third bands remain degenerate~\cite{shen_single_2010,weeks_topological_2010,jiang_lieb-like_2019}.

In the SF phase the non-zero order parameter introduces mixing between particle and hole excitations, which results in the regular anti-crossing points in the spectrum (marked with triangles in Fig.~\ref{fig:exc-spectra-smallk}). These anti-crossings appear between modes originating from the same non-interacting bands e.g., between the first and the second flat bands ($f_1$ and $f_2$ in Fig.~\ref{fig:exc-spectra-smallk}). We observe also additional anti-crossing points (marked with ellipses in Fig.~\ref{fig:exc-spectra-smallk}), which are a consequence of the non-uniform SF ground state. As opposed to the regular anti-crossing points, these appear between modes originating from different non-interacting bands. For example, the anti-crossing for~$\mu/U <0.5$ occurs between the modes originating from the~$h_3$ and~$p_1$ bands [analogous anti-crossing, but for different wavevector~$\K$, is marked with a circle in Fig.~\ref{fig:exc-spectra}(b)]. The other anti-crossing (for $\mu/U>0.5)$ in Fig.~\ref{fig:exc-spectra-smallk} occurs between the modes originating from the~$p_3$ and~$h_1$ bands. However, there are no anti-crossings between the dispersive bands and the flat bands. Since the wave function corresponding to the flat band is localized only at the rim sites, the flat band remains insensitive to the non-uniform ground state.

\subsection{Flatness parameter}

To characterize the phase-amplitude character of the modes we introduce the flatness parameter~\cite{krutitsky_ultracold_2016,di_liberto_particle-hole_2018} for each sublattice~$X$
\begin{align}
f^X_{\K,\lambda} =  \frac{R^X_{\K,\lambda}-I^X_{\K,\lambda}}{R^X_{\K,\lambda}+I^X_{\K,\lambda}},
\end{align}    
where~$R^X_{\K,\lambda} = U_{\K,\lambda}^X + V_{\K,\lambda}^X$ and~$I^X_{\K,\lambda} = U_{\K,\lambda}^X - V_{\K,\lambda}^X$ are the real and imaginary part of the order parameter oscillations~(\ref{eq:order-param-fluct}), respectively. The flatness is defined for each sublattice separately, because the order parameter as well as its oscillations take different values on each sublattice.

Flatness parameter refers to the elliptic trajectory of the order parameter around its ground state value in the complex plane. The flatness parameter takes values in the range~$[-1,1]$. The maximal value $f^X_{\K,\lambda} = 1$ corresponds to the pure amplitude character of the excitation, where the imaginary part~$I^X_{\K,\lambda}$ vanishes and the ellipse turns into a flat, horizontal line. Excitation with $f=1$ is called the Higgs mode. On the other hand, pure phase oscillations occur when the real part~$R^X_{\K,\lambda}=0$, which leads to~$f^X_{\K,\lambda} = -1$ and the ellipse turns into a flat, vertical line. When~$f^X_{\K,\lambda} >0$, the mode is more amplitude-like, while for~$f^X_{\K,\lambda} <0$ the mode is more phase-like~\cite{krutitsky_ultracold_2016,di_liberto_particle-hole_2018}. In the MI phase the order parameter fluctuations are non-zero, however the flatness is always zero. In the SF phase the first mode exhibits strong phase character corresponding to the Goldstone mode, similarly as in the simple lattices.

\begin{figure}[h]
\begin{centering}
\includegraphics[width=\columnwidth]{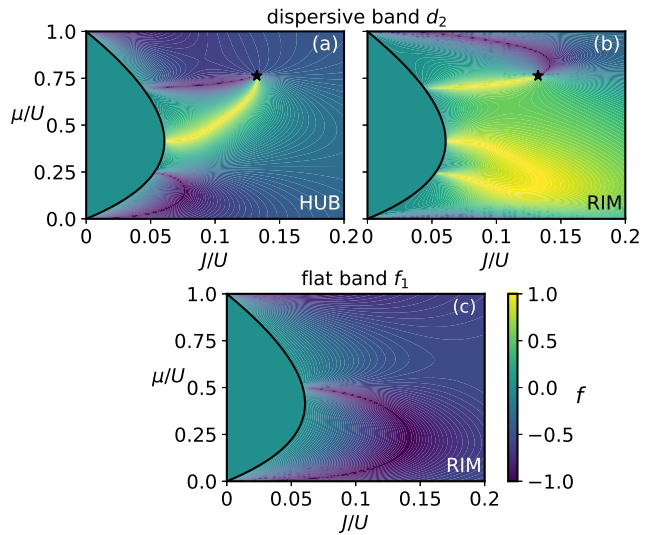} 
\par\end{centering}
\vspace{-3mm}
\caption{(Color online) Flatness parameter~$f$ as a function of the BH model parameters~$\mu/U$ and~$t/U$ for small~$\K = (\pi/100, \pi/100)$. The black line indicates the critical line of the SF-MI phase transition. Figs.~(a) and (b) presents the flatness for the dispersive band~$d_2$ (see Fig.~\ref{fig:exc-spectra-smallk}) on the hub (A) and rim (B) sublattice, respectively. Fig.~(c) presents the flatness for the flat band~$f_1$ on the rim sublattice. The flatness on the hub sublattice cannot be determined, because of the vanishing wavefunction amplitude.}
\label{fig:flatness} 
\end{figure}

In Fig.~\ref{fig:flatness} the flatness parameter for the flat ($f_1$) and the dispersive ($d_2$) mode (labeled accordingly in Fig.~\ref{fig:exc-spectra-smallk}) is presented. We have chosen~$\K = (\pi/100, \pi/100)$ similarly as in~\cite{di_liberto_particle-hole_2018}. Since for the flat band the amplitude of the wavefunction is zero on the hub (A) sublattice, the flatness parameter cannot be defined there.

For dispersive bands we do not observe an arc-shaped line of the Higgs mode, which is present in the simple lattices~\cite{di_liberto_particle-hole_2018}. This is because the curves $f=1$ do not coincide for the hub and the rim sublattice and thus there are no simultaneous pure amplitude excitations on both sublattices, except for the tips of the Mott lobes. Pure amplitude excitation is present also for the second flat band~$f_2$ (not shown), however in this case there are no order parameter oscillations on the hub sublattice.

\subsection{Particle-hole character}

The excitations can also be characterized by their particle or hole character
\begin{align}
C_{\K,\lambda} = \frac{|U_{\K,\lambda}| - |V_{\K,\lambda}|}{|U_{\K,\lambda}| + |V_{\K,\lambda}|},
\end{align}
where 
\begin{align}
|U_{\K,\lambda}| &= \sqrt{|U^A_{\K,\lambda}|^2 + |U^B_{\K,\lambda}|^2 + |U^C_{\K,\lambda}|^2}, \\
|V_{\K,\lambda}| &= \sqrt{|V^A_{\K,\lambda}|^2 + |V^B_{\K,\lambda}|^2 + |V^C_{\K,\lambda}|^2}.
\end{align}
The~$|U_{\K,\lambda}|^2$ and~$|V_{\K,\lambda}|^2$ are particle and hole strengths, respectively. The parameter~$C_{\K,\lambda}$ describes the relative particle and hole strength in a given mode. We define it on the elementary cell, not for each sublattice, to account for the spatially non-uniform distribution of the excitation amplitude among the sites in the elementary cell.

For~$C_{\K,\lambda}>0$ the excitation exhibits hole character, while for~$C_{\K,\lambda}<0$ -- particle character. $C_{\K,\lambda}=0$ indicates the particle-hole symmetry of the mode.
\begin{figure}[h]
\begin{centering}
\includegraphics[width=\columnwidth]{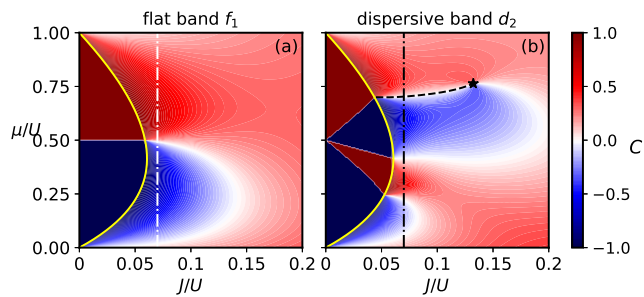} 
\par\end{centering}
\vspace{-3mm}
\caption{(Color online) Particle-hole character of the flat~$f_1$ (a) and the dispersive~$d_2$ (b) band as a function of BH model parameters~$\mu/U$ and~$t/U$ for~$\K=(\pi/100, \pi/100)$. The yellow, solid line indicates the critical line of the SF-MI phase transition. The vertical dash-dotted line indicates the line of constant~$J/U = 0.07$.  (b) The black dashed line is the particle-hole symmetry line which corresponds to pure phase- and pure amplitude oscillations in Figs.~\ref{fig:flatness}(a) and~(b), respectively.}
\label{fig:p-h-character} 
\end{figure}

In Fig.~\ref{fig:p-h-character} the parameter~$C_{\K,\lambda}$ is presented for the flat ($f_1$) and the dispersive ($d_2$) band, for~$\K=(\pi/100, \pi/100)$. States in the flat band are highly localized at the rim sites (B and C), which makes them insensitive to changes of the kinetic energy. Thus  in the MI phase the particle-hole symmetry line remains at half-integer value of the chemical potential~$\mu/U$ regardless of the hopping~$J$ [Fig.~\ref{fig:p-h-character}(a)]. In the SF phase the non-zero order parameters cause mixing of the flat bands from different Fock subspaces (different $n$), which leads to an arc-shaped particle-hole symmetry line for~$\mu/U <0.5$.

In Fig.~\ref{fig:p-h-character}(b) in the Mott insulator phase we observe four regions with alternating particle-hole character. As~$\mu/U$ increases, these regions correspond to $h_3$, $p_1$, $h_1$, and $p_3$ bands [see Fig.~\ref{fig:exc-spectra}(a)]. The lines between these regions indicate the crossings between the aforementioned bands. In the superfluid phase these lines become particle-hole symmetry lines accompanied by the anti-crossing between the~$d_2$ and~$d_3$ modes. As expected in the weak-coupling limit all excitations exhibit a weak particle character.

Generally, the lines of the particle-hole symmetry $C_{\K,\lambda}=0$ coincide with neither pure phase nor pure amplitude oscillations of the order parameter. The situation is different for the flat band, where the particle-hole symmetry line [see Fig.~\ref{fig:p-h-character}(a)] coincides with the line of the pure phase oscillations on the rim sublattice, $f_B=-1$ [see Fig.~\ref{fig:flatness}(c)]. This is due to the lack of contribution from the hub sublattice, as the amplitude of the wavefunction vanishes there. Another exception is a unique particle-hole symmetry line $C_{\K,d_2}=0$ [dashed, black line in Fig.~\ref{fig:p-h-character}(b)], which corresponds to the line of pure phase oscillations on the hub sites~[Fig.~\ref{fig:flatness}(a)] and the line of pure amplitude oscillations on the rim sites~[Fig.~\ref{fig:flatness}(b)]. For the BH model parameters on this line we can observe the anti-crossing of the $d_2$ and $d_3$ mode. This line starts at a point on the critical line, where the third particle ($p_3$) band and the first hole ($h_1$) band exchange, and terminates at the point of emergent~$d_2$ and~$d_3$ band crossing [marked with star in Figs.~\ref{fig:flatness}(a), (b), and \ref{fig:p-h-character}(b)]. The band crossing appears despite the band mixing mechanism described at the beginning of Sec.~\ref{sec:exc-spectra}. For~$\K = 0$ there is a quadratic band touching, while for higher values of~$\K$ -- a nodal-line. The point marked with star exhibits properties similar to the Lorentz-invariant points in the phase diagram~\cite{di_liberto_particle-hole_2018}, namely lines of pure phase and pure amplitude oscillations from~$d_2$ and~$d_3$ modes meet at this point. It is worth mentioning that a simplified Gutzwiller ansatz, which restricts the occupation number to~$n = n_0, n_0 \pm 1$ and thus corresponds to a projection to spin-1 model~\cite{huber_dynamical_2007}, is not able to describe this point. From this we conclude that the stabilization of this point requires at least two-particle processes. Summarizing, for carefully chosen system parameters both the pure phase and the pure amplitude oscillations can be simultaneously present within the same mode. However, they are spatially separated with pure phase oscillations existing on the hub sites and pure amplitude oscillations existing on the rim sites.

\subsection{Particle- and hole superfluid}

Near the SF-MI phase boundary two types of superfluid can be distinguished -- particle- and hole superfluid. They are found in the regions where the chemical potential~$\mu$ is a decreasing or increasing function of the hopping integral~$J$ for constant filling~$\<n\>$, respectively~\cite{wu_vortex_2004,krutitsky_dark_2010,krutitsky_ultracold_2016}. In the case of the Lieb lattice we assume the constant filling~$\<n\> = \frac{1}{3} \sum_X \<n_X\>$ per elementary cell, which is equivalent to a constant filling in the entire lattice. The condition for hole SF can be expressed as 
$\frac{d\mu}{dJ}|_{\<n\>=\text{const}} > 0$. This condition can be fulfilled only for the average occupation $\<n\>$ in range $(n_0-0.5, n_0)$. Beyond this region there is a particle superfluid, where $\frac{d\mu}{dJ}|_{\<n\>=\text{const}} < 0$. 

Experimentally, determining if the superfluid is particle- or hole dominated requires advanced techniques e.g., two-pulse Bragg spectroscopy~\cite{brunello_how_2000,vogels_experimental_2002} or ARPES-like methods~\cite{dao_probing_2009,stewart_using_2008}, which allow of the measurement of single-particle Green's function.

Here we propose a method of distinguishing between the particle and hole superfluid based on the difference of the occupation of the hub and rim sublattice $\<n_A\> - \<n_B\>$. This method would require measurements of the on-site occupations, which can be performed via high-resolution fluorescence imaging~\cite{bloch_quantum_2012}.
\begin{figure}[h]
\begin{centering}
\includegraphics[width=\columnwidth]{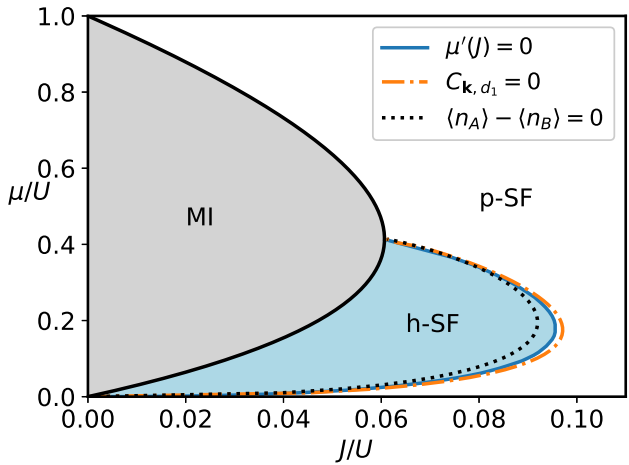} 
\par\end{centering}
\vspace{-3mm}
\caption{(Color online) The phase diagram of the BH model in the vicinity of the first Mott lobe. The hole- and particle superfluid region are separated by the blue, solid line determined from the condition for maximum of the function~$\mu(J)$ at constant filling~$\<n\>$. The condition of equal densities on the hub and the rim sites underestimates the hole SF region (black, dotted line). The orange, dash-dotted line is the particle-hole symmetry line for the Goldstone mode.}
\label{fig:p-h-symmetry} 
\end{figure}
For the Lieb lattice in the region of the hole superfluid, the average particle number on the hub sublattice~$\<n_A\>$ is lower than on the rim sublattice~$\<n_{B(C)}\>$, because higher hole-superfluid density on the hub sites leads to a stronger depletion of the average particle number. On the other hand, in the particle superfluid a higher superfluid density on the hub sites leads to a higher average occupation than on the rim sites. It is in agreement with the effect observed for soliton solutions and vortex configuration in the Bose-Hubbard model~\cite{wu_vortex_2004,krutitsky_dark_2010}, where the on-site density was shown to be generally an increasing (decreasing) function of~$|\psi_l|^2$ in the particle (hole) SF region. The line, where the average particle numbers on the hub and rim sublattices are equal, could be an indicator of a cross-over between the hole- and the particle superfluid.

In Fig.~\ref{fig:p-h-symmetry} we have presented the line determined from the condition~$\<n_A\> - \<n_B\> = 0$ (black dotted line) in the vicinity of the $\<n\> = 1$ Mott lobe as well as the line given by the condition $\frac{d\mu}{dJ}|_{\<n\>=\text{const}} = 0$ (blue solid line), which separates the hole- and particle SF regions.  We note that the agreement between these two lines is not perfect. The former underestimates the hole superfluid region and can be treated only as a guide for searching of the dark soliton solutions~\cite{krutitsky_dark_2010} and vortex configurations with maximal density at the core~\cite{wu_vortex_2004}. Because the Goldstone mode inherits the particle-hole character of the groundstate, the particle-hole symmetry line~$C_{\K\approx 0,d_1} = 0$ (orange dash-dotted line in Fig.~\ref{fig:p-h-symmetry}) is closely related to the boundary between the hole- and particle dominated superfluid. However, in contrast to the Mott insulator there are neither pure particle nor pure hole excitations in the Goldstone mode~\cite{di_liberto_particle-hole_2018}.

\section{Summary}
\label{sec:summary}

In this paper we have obtained the excitation spectra of strongly interacting bosons in the Lieb lattice, which features a bulk flat band. We have employed the time-dependent Gutzwiller mean field approach to the Bose-Hubbard model. In the MI phase the excitation spectra display behavior similar to those of simple lattices, i.e., gapped hole and particle bands. Due to a presence of three lattice sites in the elementary cell the spectrum is divided into groups of three bands. As expected, in the SF phase the lowest excitation is the gapless Goldstone mode, however the behavior of higher bands is influenced by the spatially non-uniform ground state. This is manifested in the band splitting at the corner of the BZ as well as additional avoided crossings of the dispersive bands. The flat band is not sensitive to this effect.

We have analyzed the flatness parameter for the first flat and the second dispersive modes on both sublattices. Except for the tip of the Mott lobe and the second flat band we find no pure amplitude (Higgs) mode. A unique line of particle-hole symmetry is found for the anti-crossing points of the second and third dispersive modes. On this line we can observe simultaneous and spatially separated pure phase oscillations and pure amplitude oscillations within one mode ($\lambda=d_2$), with the former localized on the hub sites and the latter on the rim sites. This line ends when the two bands ($d_2$ and $d_3$) intersect despite the non-uniform ground state. At this end point also other lines of pure phase- and pure amplitude oscillations meet.

Finally, we propose a simple method to distinguish between the hole- and particle SF for the Bose-Hubbard model in the Lieb lattice. This method relies on the fact that the order parameter in the Lieb lattice takes greater value on the hub ($A$) sublattice. This in turn leads to a lower average density on~$A$ sites than on~$B$ or $C$ sites in the hole superfluid regime. Equal densities on~$A$ and on~$B$ (or $C$) sites signals the crossover between the hole- and particle SF. This condition reproduces the boundary between these two types of superfluid~\cite{wu_vortex_2004,krutitsky_dark_2010,krutitsky_ultracold_2016} only approximately, however the proposed method could be easily applied in an experiment as it requires only in-situ measurement of the local atomic density.


\begin{acknowledgments}

The authors would like to thank T.~A.~Zaleski for useful discussions and careful reading of the manuscript.

\end{acknowledgments}

\section*{Author contributions}
Both authors contributed equally to this work.


\begin{widetext}
\appendix*
\section{Generalized Bogoliubov equations}

\setcounter{equation}{0}
\renewcommand{\theequation}{A\arabic{equation}}

\label{sec:appA}

The first order expansion in the perturbation~$\delta c$~(\ref{eq:perturbation}) of the Gutzwiller equations~(\ref{eq:Gutzwiller_eqs}) has the following form:
\begin{subequations}
\label{eq:u-Bogoliubov-like}
\begin{align}
\hbar \omega_\K \, u_{\K,n}^A &= \varepsilon_n^A u_{\K,n}^A -2J (\psi_B+\psi_C) \( \sqrt{n} \, u_{\K,n-1}^A + \sqrt{n+1} \, u_{\K,n+1}^A \) \nonumber  \\
&+ \sum_{\np} \[ \mathcal{J}_y \( S_{n\np}^{AB} u_{\K,\np}^B + D_{n\np}^{AB} v_{\K,\np}^B \) + \mathcal{J}_x \( S_{n\np}^{AC} u_{\K,\np}^C + D_{n\np}^{AC} v_{\K,\np}^C \) \], \\
\hbar \omega_\K \, u_{\K,n}^B &= \varepsilon_n^B u_{\K,n}^B -2 J \psi_A \( \sqrt{n} \, u_{\K,n-1}^B + \sqrt{n+1} \, u_{\K,n+1}^B \)+ \sum_{\np} \mathcal{J}_y \( S_{n\np}^{BA} u_{\K,\np}^A + D_{n\np}^{BA} v_{\K,\np}^A \), \\
\hbar \omega_\K \, u_{\K,n}^C &= \varepsilon_n^C u_{\K,n}^C -2 J \psi_A \( \sqrt{n} \, u_{\K,n-1}^C + \sqrt{n+1} \, u_{\K,n+1}^C \) + \sum_{\np} \mathcal{J}_x \( S_{n\np}^{CA} u_{\K,\np}^A + D_{n\np}^{CA} v_{\K,\np}^A \),  
\end{align}
\end{subequations}
\begin{subequations}
\label{eq:v-Bogoliubov-like}
\begin{align}
\hbar \omega_\K \, v_{\K,n}^A &= -\varepsilon_n^A v_{\K,n}^A + 2 J (\psi_B+\psi_C) \( \sqrt{n} \, v_{\K,n-1}^A + \sqrt{n+1} \, v_{\K,n+1}^A \) \nonumber  \\
&- \sum_{\np} \[ \mathcal{J}_y \( S_{n\np}^{AB} \, v_{\K,\np}^B + D_{n\np}^{AB} \, u_{\K,\np}^B \) + \mathcal{J}_x \( S_{n\np}^{AC} \, v_{\K,\np}^C + D_{n\np}^{AC} \, u_{\K,\np}^C \) \], \\
\hbar \omega_\K \, v_{\K,n}^B &= -\varepsilon_n^B v_{\K,n}^B +2J \psi_A \( \sqrt{n} \, v_{\K,n-1}^B + \sqrt{n+1} \, v_{\K,n+1}^B \) - \sum_{\np} \mathcal{J}_y \( S_{n\np}^{BA} \, v_{\K,\np}^A + D_{n\np}^{BA} \, u_{\K,\np}^A \), \\
\hbar \omega_\K \, v_{\K,n}^C &= -\varepsilon_n^C v_{\K,n}^C +2 J \psi_A \( \sqrt{n} \, v_{\K,n-1}^C + \sqrt{n+1} \, v_{\K,n+1}^C \) - \sum_{\np} \mathcal{J}_x \( S_{n\np}^{CA} \, v_{\K,\np}^A + D_{n\np}^{CA} \, u_{\K,\np}^A \).  
\end{align}
\end{subequations}
Without a loss of generality we choose the coefficients~$c_n^X$, $u_{\K,n}^X$, and $v_{\K,n}^X$, where~$X=A,B,C$, to be real~\cite{krutitsky_ultracold_2016}. The energy~$\varepsilon_n^X = E_n - \hbar\omega_0^X$ and $\psi_X = \sum_{n=1}^\infty \sqrt{n} \, c_{n-1}^X c_n^X$ is the ground state order parameter on the~$X$ site. The couplings between the sublattice~$A$ and $B$, and $A$ and $C$ are given by
\begin{align}
\mathcal{J}_x \equiv \mathcal{J}_x(\K) &= -2 J \cos\( \frac{k_x \mathrm{a}}{2} \), \\
\mathcal{J}_y \equiv \mathcal{J}_y(\K) &= -2 J \cos\( \frac{k_y \mathrm{a}}{2} \),
\end{align} 
respectively. The lattice spacing is denoted as~'$\mathrm{a}$' [see Fig.~\ref{fig:lieb-lattice}(a)]. Functions~$S_{n\np}^{XY}$ and $D_{n\np}^{XY}$ ($X,Y=A,B,C$) couple coefficients with $n$ and $\np$ occupation numbers and are defined as
\begin{align}
S_{n\np}^{XY} &= \sqrt{n} \sqrt{\np} \, c_{n-1}^X \, c_{\np-1}^Y + \sqrt{n+1} \, \sqrt{\np+1} \, c_{n+1}^X \, c_{\np+1}^Y, \\
D_{n\np}^{XY} &= \sqrt{n+1} \sqrt{\np} \, c_{n+1}^X \, c_{\np-1}^Y + \sqrt{n} \, \sqrt{\np+1} \, c_{n-1}^X \, c_{\np+1}^Y .
\end{align}
The Bogoliubov-like equations~(\ref{eq:u-Bogoliubov-like}) and~(\ref{eq:v-Bogoliubov-like}) can be rewritten in the matrix form [see Eq.~(\ref{eq:Bogoliubov-eqs})]
\begin{align}
\hbar \omega_\K 
\begin{bmatrix}
\vec{u}_\K \\
\vec{v}_\K
\end{bmatrix} = 
\begin{bmatrix}
M_\K & N_\K \\
-N_\K & -M_\K
\end{bmatrix}
\begin{bmatrix}
\vec{u}_\K \\
\vec{v}_\K
\end{bmatrix},
\end{align}
where
\begin{align}
\begin{bmatrix} \vec{u}^\top_\K,\vec{v}^\top_\K \end{bmatrix} = \begin{bmatrix} u^A_{\K,0}, u^B_{\K,0}, u^C_{\K,0}, u^A_{\K,1}, u^B_{\K,1}, u^C_{\K,1} \dots ; v^A_{\K,0}, v^B_{\K,0}, v^C_{\K,0}, v^A_{\K,1}, v^B_{\K,1}, v^C_{\K,1}, \dots \end{bmatrix}
\end{align}
The~$M_\K$ matrix is given as
\begin{align}
M_\K = \begin{bmatrix} 
\varepsilon_0^A & \mathcal{J}_y S_{00}^{AB} & \mathcal{J}_x S_{00}^{AC} & \mathcal{J}_{BC} & \mathcal{J}_y S_{01}^{AB} & \mathcal{J}_x S_{01}^{AC} & 0 & \mathcal{J}_y S_{02}^{AB} & \mathcal{J}_x S_{02}^{AC} & \cdots \\
\mathcal{J}_y S_{00}^{BA} & \varepsilon_0^B & 0 & \mathcal{J}_y S_{01}^{BA} & \mathcal{J}_A & 0  & \mathcal{J}_y S_{02}^{BA} & 0 & 0 & \cdots \\
\mathcal{J}_x S_{00}^{CA} & 0 & \varepsilon_0^C & \mathcal{J}_x S_{01}^{CA} & 0 & \mathcal{J}_A & \mathcal{J}_x S_{02}^{CA} & 0 & 0 & \cdots \\
\mathcal{J}_{BC} & \mathcal{J}_y S_{10}^{AB} & \mathcal{J}_x S_{10}^{AC} & \varepsilon_1^A & \mathcal{J}_y S_{11}^{AB} & \mathcal{J}_x S_{11}^{AC} & \mathcal{J}_{BC} \sqrt{2} & \mathcal{J}_y S_{12}^{AB} & \mathcal{J}_x S_{12}^{AC}  & \cdots \\
\mathcal{J}_y S_{10}^{BA} & \mathcal{J}_A & 0 & \mathcal{J}_y S_{11}^{BA} & \varepsilon_1^B & 0 &  \mathcal{J}_y S_{12}^{BA} & \mathcal{J}_A \sqrt{2} & 0 & \cdots \\
\mathcal{J}_x S_{10}^{CA} & 0 & \mathcal{J}_A & \mathcal{J}_x S_{11}^{CA} & 0 & \varepsilon_1^C & \mathcal{J}_x S_{12}^{CA} & 0 & \mathcal{J}_A \sqrt{2} & \cdots \\
0 & \mathcal{J}_y S_{20}^{AB} & \mathcal{J}_x S_{20}^{AC} & \mathcal{J}_{BC}\sqrt{2} & \mathcal{J}_y S_{21}^{AB} & \mathcal{J}_x S_{21}^{AC} & \varepsilon_2^A & \mathcal{J}_y S_{22}^{AB} & \mathcal{J}_x S_{22}^{AC} & \cdots \\
\mathcal{J}_y S_{20}^{BA} & 0 & 0 & \mathcal{J}_y S_{21}^{BA} & \mathcal{J}_A\sqrt{2} & 0 & \mathcal{J}_y S_{22}^{BA} & \varepsilon_2^B & 0  & \cdots \\
\mathcal{J}_x S_{20}^{CA} & 0 & 0 & \mathcal{J}_x S_{21}^{CA} & 0 & \mathcal{J}_A\sqrt{2} & \mathcal{J}_x S_{22}^{CA} & 0 & \varepsilon_2^C & \cdots \\
\vdots & \vdots & \vdots & \vdots & \vdots & \vdots & \vdots & \vdots & \vdots & \ddots 
\end{bmatrix},
\end{align}
where
\begin{align}
\mathcal{J}_{BC} &\equiv -2 J (\psi_B + \psi_C), \\
\mathcal{J}_A &\equiv -2 J \psi_A.
\end{align}
Matrix~$N_\K$ can be represented in a block form
\begin{align}
N_\K = \begin{bmatrix}
N_{00} & N_{01} & N_{02} & \cdots \\
N_{10} & N_{11} & N_{12} & \cdots \\
N_{20} & N_{21} & N_{22} & \cdots \\
\vdots & \vdots & \vdots & \ddots 
\end{bmatrix},
\end{align}
where 
\begin{align}
N_{n\np} = \begin{bmatrix}
0 & \mathcal{J}_y D_{n\np}^{AB} & \mathcal{J}_x D_{n\np}^{AC} \\
\mathcal{J}_y D_{n\np}^{BA} & 0 & 0 \\
\mathcal{J}_x D_{n\np}^{CA} & 0 & 0
\end{bmatrix}.
\end{align}

\end{widetext}

\bibliographystyle{apsrev}
\bibliography{lieb-bibliography}


\end{document}